# Dispersion and damping of ion-acoustic waves in the plasma with a regularized kappa-distribution

Huo Rui, Du Jiulin

*Department of Physics, School of Science, Tianjin University, Tianjin 300350, China*

**Abstract** The dispersion and damping of ion-acoustic waves in the plasma with a regularized kappa-distribution are studied. The generalized dispersion relation and damping rate are derived, which both depend significantly on the parameters $\alpha$ and $\kappa$. The numerical analyses show that the wave frequency ($\omega_r/\omega_{pi}$) and the damping rate ($-\gamma/\omega_r$) of ion-acoustic waves in the plasma with the regularized kappa-distribution are both generally less than those in the plasma with the kappa-distribution, and if $\kappa < 3/2$ the ion-acoustic waves and their damping rate exist in the plasma with the regularized kappa-distribution.

## 1. Introduction

Ion-acoustic waves are basic low frequency oscillation waves, which can be observed in space and laboratory plasmas. The kinetic model is a basic theoretical study of ion-acoustic wave in plasmas, it is a statistical model based on the kinetic equations and the distribution functions of plasma components. Usually, one assumes the plasma components to be a Maxwellian distribution and studies the dispersion and damping rate of ion-acoustic waves from the kinetic model [1]. However, for the systems with long-range interaction, long memory or multifractal structure, the velocity distribution of particles is not a Maxwellian one, but often has a long power-law tail, a non-Maxwell distribution. Non-Maxwellian distributions are very common in nonequilibrium astrophysical and space plasmas, such as the $q$-distribution in nonextensive statistics [2], the kappa-distribution [3], the non-thermal $\alpha$-distribution [4], the Vasyliunas–Cairns distribution [5], and so on. The plasma dispersion function was modified when the plasma is a kappa-distribution [6]. The $\alpha$-distribution was used to simulate the non-thermal electrons observed by Freja and Viking satellites in the space plasma [7]. The ion-acoustic waves of the plasmas with the power-law $q$-distribution were studied in the framework of nonextensive statistics [8, 9]. The Landau damping of electrostatic modes with a $\alpha$-distribution was presented [10]. The ion-acoustic waves in the plasma with a non-thermal Vasyliunas–Cairns distribution was also studied [11].

The kappa-distribution and the related property have been widely applied to describe the collective behaviors of the plasmas such as the solar wind [12], the flare [13], the Earth magnetopause and magnetosheath [14], the interstellar medium and some planetary magnetospheres [15], etc. But there might be a mathematical issue in the kappa-distribution function due to κ > 3/2, i.e., existence of the velocity moments is restricted. Recently, a regularized kappa-distribution function was defined [16], with which the exponential cutoff of high energy tail made all velocity moments valid for all $\kappa > 0$. If one used the regularized kappa-distribution to replace the standard kappa-distribution, the properties of plasmas might not be the same. For example, it was employed to study moments in magneto-hydrodynamics [17], solitary ion-acoustic waves [18], and Langmuir waves [19]. In this paper, we study the dispersion and damping rate of ion-acoustic waves in the plasma with the regularized kappa-distribution.

The paper is organized as follows. In section 2, the dispersion relation and the damping rate of ion-acoustic waves are derived from the kinetic theory. In section 3, numerical analyses are made. And in section 4, the conclusion is given.

## 2. The dispersion and damping of ion-acoustic waves

We now introduce the theory of ion-acoustic waves for the unmagnetized and collisionless plasma. The kinetic model is the Vlasov equation [20],

$$\frac{\partial f_s}{\partial t} + \mathbf{v} \cdot \frac{\partial f_s}{\partial \mathbf{r}} + \frac{q_s}{m_s}(\mathbf{E} + \mathbf{v} \times \mathbf{B}) \cdot \frac{\partial f_s}{\partial \mathbf{v}} = 0, \tag{1}$$

where $f_s$ is the distribution function of the plasma components ($s = i$ for ions and $s = e$ for electrons); $m_s$ and $q_s$ are respectively mass and charge of a particle. If there is a weak disturbance around a state, the linear perturbation in field $\mathbf{E}$, $\mathbf{B}$ and distribution function was considered as

$$f_s(\mathbf{r},\mathbf{v},t) = f_{s0}(\mathbf{r},\mathbf{v},t) + f_{s1}(\mathbf{r},\mathbf{v},t),$$
$$\mathbf{E}(\mathbf{r},t) = \mathbf{E}_0(\mathbf{r},t) + \mathbf{E}_1(\mathbf{r},t),$$
$$\mathbf{B}(\mathbf{r},t) = \mathbf{B}_0(\mathbf{r},t) + \mathbf{B}_1(\mathbf{r},t). \tag{2}$$

But, in the case of electrostatic oscillation, it is unnecessary to consider the magnetic field, and the disturbing electric field is only electrostatic, i.e.,

$$\mathbf{E}_0 = \mathbf{B}_0 = 0 \qquad \mathbf{E}_1 = -\nabla \varphi. \tag{3}$$

We assume that $f_{s0}$ and $\mathbf{E}_0$ are known and satisfy the Vlasov equation and Poisson equation:

$$\frac{\partial f_{s0}}{\partial t} + \mathbf{v} \cdot \frac{\partial f_{s0}}{\partial \mathbf{r}} + \frac{q_s}{m_s}\mathbf{E}_0 \cdot \frac{\partial f_{s0}}{\partial \mathbf{v}} = 0, \tag{4}$$

$$\nabla \cdot \mathbf{E}_0 = \frac{1}{\varepsilon_0}\sum_s q_s \int f_{s0} d\mathbf{v}. \tag{5}$$

Substituting Eqs. (2)-(5) into Eq.(1), we obtain the equation,

$$\frac{\partial f_{s1}}{\partial t} + \mathbf{v} \cdot \frac{\partial f_{s1}}{\partial \mathbf{r}} - \frac{q_s \nabla \varphi}{m_s} \frac{\partial f_{s0}}{\partial \mathbf{v}} = 0, \tag{6}$$

and

$$\nabla^2 \varphi = \sum_s \nabla^2 \varphi_s = -\frac{1}{\varepsilon_0}\sum_s q_s \int f_{s1} d\mathbf{v}. \tag{7}$$

If frequency and wave vector is $\omega$ and $\mathbf{k}$, and, without lose of generality, the x-axis is along the direction of wave vector, then the solution of $f_{s1}$ and $\varphi$ has the following form:

$$f_{s1}(\mathbf{x},\mathbf{v},t) = f_{sk}(\mathbf{v})\exp(i\mathbf{k}\cdot\mathbf{x})\exp(-i\omega t), \quad \varphi(\mathbf{x},t) = \varphi_k \exp(i\mathbf{k}\cdot\mathbf{x})\exp(-i\omega t). \tag{8}$$

From Eq. (6) and Eq.(7) we can get that

$$f_{sk} = -\frac{(q_s/m_s)(\mathbf{k}\cdot\nabla_\mathbf{v} f_{s0})}{\omega - \mathbf{k}\cdot\mathbf{v}}\varphi_k, \tag{9}$$

and

$$k^2\varphi_k(1 + \sum_s \frac{q_s^2 n_{s0}}{\varepsilon_0 m_s}\frac{1}{k^2}\int_{-\infty}^{\infty}\frac{\partial \hat{f}_{s0}/\partial v_x}{\omega/k - v_x}dv_x) = 0. \tag{10}$$

Therefore, the dispersion function of ion-acoustic waves is obtained from the equation [20],

$$D(k,\omega) = 1 + \sum_s \frac{\omega_{ps}^2}{k^2}\int_{-\infty}^{\infty}\frac{\partial \hat{f}_{s0}/\partial v_x}{\omega/k - v_x}dv_x = 0, \tag{11}$$

where $\hat{f}_{s0}$ is the normalized distribution function, $\hat{f}_{s0} = f_{s0}/n_{s0}$, and $\omega_{ps}^2 = \frac{q_s^2 n_{s0}}{\varepsilon_0 m_s}$ is the natural oscillation plasma frequency.

From Eq.(11) we see that the dispersion relation of ion-acoustic waves is determined by the velocity distribution function of particles in the plasma. Thus if one used the regularized kappa-distribution to replace the standard kappa-distribution, the property of the ion-acoustic waves would be not the same. The regularized kappa-distribution was defined by introducing a parameter α > 0 [16] by

$$f_{s0}(v) = n_{s0}N_{\alpha\kappa}\left(1 + \frac{v^2}{\kappa v_{Ts}^2}\right)^{-\kappa-1}\exp\left(-\alpha^2\frac{v^2}{v_{Ts}^2}\right), \tag{12}$$

where $N_{\alpha\kappa} = (\pi\kappa v_{Ts}^2)^{-3/2}U^{-1}(3/2, 3/2-\kappa; \alpha^2\kappa)$ is the normalized constant [19], $v_{Ts} = \sqrt{k_B T_s/m_s}$ is the thermal speed, κ>0 and α ≥0 are two non-thermal parameters. $U(a, b; z)$ in $N_{\alpha\kappa}$ is a Tricomi function (or a Kummer U function), defined by

$$U(a,b;z) = \frac{1}{\Gamma(a)}\int_0^\infty \exp(-zt)t^{a-1}(1+t)^{b-a-1}dt, \tag{13}$$

where Γ (a) is a Gamma function. When we take α=0, (12) becomes the kappa-distribution, and when we take α=0 and κ→∞, (12) becomes to a Maxwellian distribution.

By integrating the function (12) over $v_y$ and $v_z$, one can obtain the 1-dimensional expression of the regularized kappa-distribution [18,19],

$$f_{s0}(v_x) = n_{s0}N_{\alpha\kappa}\pi\kappa v_{Ts}^2 U\left(1, 1-\kappa; \alpha^2\kappa(1+\frac{v_x^2}{\kappa v_{Ts}^2})\right)\left(1 + \frac{v_x^2}{\kappa v_{Ts}^2}\right)^{-\kappa}\exp\left(-\alpha^2\frac{v_x^2}{v_{Ts}^2}\right). \tag{14}$$

When we take α=0, the regularized kappa-distribution (14) recovers to the 1-dimensional kappa-distribution,

$$f_\kappa(v_x) = \frac{n_{s0}\Gamma(\kappa)}{\sqrt{\pi\kappa v_{ts}^2}\Gamma(\kappa-\frac{1}{2})}\left(1 + \frac{v_x^2}{\kappa v_{ts}^2}\right)^{-\kappa}, \tag{15}$$

which is equivalent to the *q*-distribution in nonextensive statistics when it is applied to study astrophysical and space plasmas [21,22].

Now using the new distribution function and substituting (14) into Eq. (11), one gets that

$$D_{\alpha,\kappa}(\omega,k) = 1 + 2\pi \sum_s \frac{\omega_{ps}^2}{k^2} N_{\alpha\kappa} \int_{-\infty}^{\infty} dv_x \left(1 + \frac{v_x^2}{\kappa v_{Ts}^2}\right)^{-1-\kappa} \frac{-v_x \exp(-\alpha^2 v_x^2 / v_{Ts}^2)}{\omega/k - v_x}$$

$$= 1 + 2\pi \sum_s \frac{\omega_{ps}^2}{k^2} N_{\alpha\kappa} \left[ \int_{-\infty}^{\infty} dv_x \left(1 + \frac{v_x^2}{\kappa v_{Ts}^2}\right)^{-1-\kappa} \exp(-\alpha^2 v_x^2 / v_{Ts}^2) - \frac{\omega}{k} \int_{-\infty}^{\infty} dv_x \left(1 + \frac{v_x^2}{\kappa v_{Ts}^2}\right)^{-1-\kappa} \frac{\exp(-\alpha^2 v_x^2 / v_{Ts}^2)}{\omega/k - v_x} \right]$$

$$= 1 + \sum_s \frac{\omega_{ps}^2}{k^2} \frac{2U(\frac{1}{2},\frac{1}{2}-\kappa;\alpha^2\kappa)}{\kappa v_{Ts}^2 U(\frac{3}{2},\frac{3}{2}-\kappa;\alpha^2\kappa)} \left[ 1 - \frac{\omega/k}{\sqrt{\pi\kappa} U(\frac{1}{2},\frac{1}{2}-\kappa;\alpha^2\kappa)} \int_{-\infty}^{\infty} d\left(\frac{v_x}{v_{Ts}}\right) \left(1 + \frac{v_x^2}{\kappa v_{Ts}^2}\right)^{-1-\kappa} \frac{\exp(-\alpha^2 v_x^2 / v_{Ts}^2)}{\omega/k - v_x} \right]$$

$$= 1 + \sum_s \frac{\omega_{ps}^2}{k^2} \frac{2}{\kappa v_{Ts}^2} \frac{U(\frac{1}{2},\frac{1}{2}-\kappa;\alpha^2\kappa)}{U(\frac{3}{2},\frac{3}{2}-\kappa;\alpha^2\kappa)} \left[1 - \xi_s Z_{\alpha,\kappa}(\xi_s)\right], \tag{16}$$

where $\xi_s = \omega/(k v_{Ts})$ is the ratio of the phase velocity $\omega/k$ to the thermal speed $v_{Ts}$, and $Z(\xi_s)$ is the generalized dispersion function,

$$Z_{\alpha,\kappa}(\xi_s) = \frac{1}{\sqrt{\pi\kappa} U(\frac{1}{2},\frac{1}{2}-\kappa;\alpha^2\kappa)} \int_{-\infty}^{\infty} dx \left(1 + \frac{x^2}{\kappa}\right)^{-\kappa-1} \frac{\exp(-\alpha^2 x^2)}{\xi_s - x}. \tag{17}$$

Obviously, there exists a singularity in (17) at $x = \xi_s$; According to Plemelj formula [20], if the frequency is written as the complex form, $\omega = \omega_r + i\gamma$, then Eq. (17) can be expressed as

$$Z_{\alpha,\kappa}(\xi_s) = \frac{1}{\sqrt{\pi\kappa} U(\frac{1}{2},\frac{1}{2}-\kappa;\alpha^2\kappa)} \Pr \int_{-\infty}^{\infty} \left(1 + \frac{x^2}{\kappa}\right)^{-\kappa-1} \frac{\exp(-\alpha^2 x^2)}{\xi_s - x} dx$$

$$- i\sqrt{\frac{\pi}{\kappa}} \frac{1}{U(\frac{1}{2},\frac{1}{2}-\kappa;\alpha^2\kappa)} \left(1 + \frac{\xi_s^2}{\kappa}\right)^{-\kappa-1} \exp(-\alpha^2 \xi_s^2). \tag{18}$$

Then Eq. (16) can be written as

$$D_{\alpha,\kappa}(\omega,k) = 1 + \sum_s \frac{\omega_{ps}^2}{k^2} \frac{2}{\kappa v_{Ts}^2} \frac{U(\frac{1}{2},\frac{1}{2}-\kappa;\alpha^2\kappa)}{U(\frac{3}{2},\frac{3}{2}-\kappa;\alpha^2\kappa)} \left[ 1 - \frac{\xi_s}{\sqrt{\pi\kappa} U(\frac{1}{2},\frac{1}{2}-\kappa;\alpha^2\kappa)} \Pr \int_{-\infty}^{\infty} dx \left(1 + \frac{x^2}{\kappa}\right)^{-\kappa-1} \frac{\exp(-\alpha^2 x^2)}{\xi_s - x} \right.$$

$$\left. - i \frac{\sqrt{\pi/\kappa}}{U(\frac{1}{2},\frac{1}{2}-\kappa;\alpha^2\kappa)} \xi_s \left(1 + \frac{\xi_s^2}{\kappa}\right)^{-\kappa-1} \exp(-\alpha^2 \xi_s^2) \right]. \tag{19}$$

For the ion-acoustic waves, the relationship between the phase velocity and the thermal velocities is usually that $v_{Ti} < \omega/k < v_{Te}$, so the modified dispersion function can be expanded in such a way with a small variable expansion ($\xi_e \ll 1$) for electrons but with a large variable expansion ($\xi_i \gg 1$) for ions. In this way, making series expansion for the integrand in Eq. (19) we have that (see Appendix),

$$D_{\alpha,k}(\omega,k) = 1 - \frac{\omega_{pi}^2}{\omega^2} - \frac{3\kappa \omega_{pi}^2 k^2 v_{Ti}^2}{2\omega^4} \frac{U(\frac{5}{2},\frac{5}{2}-\kappa;\alpha^2\kappa)}{U(\frac{3}{2},\frac{3}{2}-\kappa;\alpha^2\kappa)} + \frac{2\omega_{pe}^2}{\kappa v_{Te}^2 k^2} \frac{U(\frac{1}{2},\frac{1}{2}-\kappa;\alpha^2\kappa)}{U(\frac{3}{2},\frac{3}{2}-\kappa;\alpha^2\kappa)}$$

$$+ i\sqrt{\frac{\pi}{\kappa^3}} \frac{2\omega}{U(\frac{3}{2},\frac{3}{2}-\kappa;\alpha^2\kappa)} \sum_s \frac{\omega_{ps}^2}{k^3 v_{Ts}^3} \left(1 + \frac{\omega^2}{\kappa k^2 v_{Ts}^2}\right)^{-\kappa-1} \exp\left(\frac{-\alpha^2 \omega^2}{k^2 v_{Ts}^2}\right). \tag{20}$$

Inserting $\omega = \omega_r + i\gamma$ in Eq. (20) and, according the dispersion equation (11), making the real part of (20) zero, we obtain that

$$1 - \frac{\omega_{pi}^2}{\omega_r^2} - \frac{3k^2\omega_{pi}^2 v_{Ti}^2}{2\omega_r^4}\frac{\kappa U(\frac{5}{2},\frac{5}{2}-\kappa;\alpha^2\kappa)}{U(\frac{3}{2},\frac{3}{2}-\kappa;\alpha^2\kappa)} + \frac{2\omega_{pe}^2}{\kappa v_{Te}^2 k^2}\frac{U(\frac{1}{2},\frac{1}{2}-\kappa;\alpha^2\kappa)}{U(\frac{3}{2},\frac{3}{2}-\kappa;\alpha^2\kappa)} = 0. \quad (21)$$

Further, we use $\lambda_{Ds}^2 = v_{Ts}^2/2\omega_{ps}^2$, and write Eq.(21) as

$$\omega_r^2 = \frac{k^2\lambda_{De}^2 \omega_{pi}^2}{k^2\lambda_{De}^2 + \frac{U(\frac{1}{2},\frac{1}{2}-\kappa;\alpha^2\kappa)}{\kappa U(\frac{3}{2},\frac{3}{2}-\kappa;\alpha^2\kappa)}}\left(1 + \frac{k^2 v_{Ti}^2}{\omega_r^2}\frac{3\kappa U(\frac{5}{2},\frac{5}{2}-\kappa;\alpha^2\kappa)}{2U(\frac{3}{2},\frac{3}{2}-\kappa;\alpha^2\kappa)}\right), \quad (22)$$

Because $kv_{Ti}/\omega_r$ is very small and the second term in the above bracket is also very small, we can approximately apply this expression,

$$\omega_r^2 \approx \frac{k^2\lambda_{De}^2 \omega_{pi}^2}{k^2\lambda_{De}^2 + \frac{U(\frac{1}{2},\frac{1}{2}-\kappa;\alpha^2\kappa)}{\kappa U(\frac{3}{2},\frac{3}{2}-\kappa;\alpha^2\kappa)}}, \quad (23)$$

to the right-hand side of Eq.(22) [18], and then we derive the generalized dispersion relation,

$$\omega_r^2 = \frac{k^2 k_B T_e / m_i}{k^2\lambda_{De}^2 + \frac{U(\frac{1}{2},\frac{1}{2}-\kappa;\alpha^2\kappa)}{\kappa U(\frac{3}{2},\frac{3}{2}-\kappa;\alpha^2\kappa)}} + k^2 v_{Ti}^2 \frac{3\kappa U(\frac{5}{2},\frac{5}{2}-\kappa;\alpha^2\kappa)}{2U(\frac{3}{2},\frac{3}{2}-\kappa;\alpha^2\kappa)}, \quad (24)$$

When we take $\alpha = 0$, this expression (24) can correctly recover to the dispersion relation of the plasma with the $q$-distribution (or the $\kappa$-distribution) in nonextensive statistics [8].

For a weak Landau damping, the damping rate can be obtained [18] by the equation,

$$\gamma = -\frac{\mathrm{Im}\, D_{\alpha,\kappa}(k,\omega_r)}{\partial \mathrm{Re}\, D_{\alpha,\kappa}(k,\omega_r)/\partial \omega_r}, \quad (25)$$

where from Eq.(20) we have that

$$\mathrm{Im}\, D_{\alpha,\kappa}(k,\omega_r) = \sqrt{\frac{\pi}{\kappa^3}}\frac{2\omega_r}{U(\frac{3}{2},\frac{3}{2}-\kappa;\alpha^2\kappa)}\sum_s \frac{\omega_{ps}^2}{k^3 v_{Ts}^3}\left(1 + \frac{\omega_r^2}{\kappa k^2 v_{Ts}^2}\right)^{-\kappa-1}\exp\left(-\frac{\alpha^2 \omega_r^2}{k^2 v_{Ts}^2}\right),$$

$$\mathrm{Re}\, D_{\alpha,\kappa}(k,\omega_r) = 1 - \frac{\omega_{pi}^2}{\omega_r^2} - \frac{3k^2 v_{Ti}^2 \omega_{pi}^2}{2\omega_r^4}\frac{\kappa U(\frac{5}{2},\frac{5}{2}-\kappa;\alpha^2\kappa)}{U(\frac{3}{2},\frac{3}{2}-\kappa;\alpha^2\kappa)} + \frac{2\omega_{pe}^2}{\kappa k^2 v_{Te}^2}\frac{U(\frac{1}{2},\frac{1}{2}-\kappa;\alpha^2\kappa)}{U(\frac{3}{2},\frac{3}{2}-\kappa;\alpha^2\kappa)}.$$

And then we derive the damping rate of the ion-acoustic waves,

$$\gamma = -\frac{\sqrt{\pi/\kappa^3}}{U(\frac{3}{2},\frac{3}{2}-\kappa;\alpha^2\kappa)}\frac{\omega_r^4}{k^3 v_{Ti}^3}\left[\left(1 + \frac{\omega_r^2}{\kappa k^2 v_{Ti}^2}\right)^{-\kappa-1}\exp\left(-\frac{\alpha^2 \omega_r^2}{k^2 v_{Ti}^2}\right)\right.$$

$$\left. + \frac{\omega_{pe}^2}{\omega_{pi}^2}\frac{v_{Ti}^3}{v_{Te}^3}\left(1 + \frac{\omega_r^2}{\kappa k^2 v_{Te}^2}\right)^{-\kappa-1}\exp\left(-\frac{\alpha^2 \omega_r^2}{k^2 v_{Te}^2}\right)\right]. \quad (26)$$

In order to do numerical analyses, we need to express Eq.(26) as a function of temperatures and masses of the plasma components. By using Eq.(15), we have that

$$\frac{\omega_r^2}{k^2 v_{Ti}^2} = \frac{1}{k^2 \lambda_{De}^2 + \frac{U(\frac{1}{2},\frac{1}{2}-\kappa;\alpha^2\kappa)}{\kappa U(\frac{3}{2},\frac{3}{2}-\kappa;\alpha^2\kappa)}} \frac{T_e}{2T_i} + \frac{3\kappa U(\frac{5}{2},\frac{5}{2}-\kappa;\alpha^2\kappa)}{2U(\frac{3}{2},\frac{3}{2}-\kappa;\alpha^2\kappa)}, \tag{27}$$

$$\frac{\omega_r^2}{k^2 v_{Te}^2} = \frac{\omega_r^2}{k^2 v_{Ti}^2} \frac{m_e T_i}{m_i T_e} = \frac{m_e}{2m_i}\left(\frac{1}{k^2 \lambda_{De}^2 + \frac{U(\frac{1}{2},\frac{1}{2}-\kappa;\alpha^2\kappa)}{\kappa U(\frac{3}{2},\frac{3}{2}-\kappa;\alpha^2\kappa)}} + \frac{T_i}{T_e}\frac{3\kappa U(\frac{5}{2},\frac{5}{2}-\kappa;\alpha^2\kappa)}{U(\frac{3}{2},\frac{3}{2}-\kappa;\alpha^2\kappa)}\right). \tag{28}$$

Substituting Eq.(27), Eq.(28) and $\frac{v_{Ti}^3}{v_{Te}^3}\frac{\omega_{pe}^2}{\omega_{pi}^2} = (T_i/T_e)^{3/2}\sqrt{m_e/m_i}$ into Eq.(26), it becomes

$$\frac{\gamma}{\omega_r} = -\frac{\sqrt{\pi/\kappa^3}\chi^3}{U(\frac{3}{2},\frac{3}{2}-\kappa;\alpha^2\kappa)}\left[\left(1+\frac{\chi^2}{\kappa}\right)^{-\kappa-1}\exp(-\alpha^2\chi^2) + \left(\frac{T_i}{T_e}\right)^{3/2}\sqrt{\frac{m_e}{m_i}}\left(1+\frac{\zeta^2}{\kappa}\right)^{-\kappa-1}\exp(-\alpha^2\zeta^2)\right], \tag{29}$$

where we have denoted that

$$\chi \equiv \frac{\omega_r}{kv_{Ti}} = \left(\frac{1}{k^2 \lambda_{De}^2 + \frac{U(\frac{1}{2},\frac{1}{2}-\kappa;\alpha^2\kappa)}{\kappa U(\frac{3}{2},\frac{3}{2}-\kappa;\alpha^2\kappa)}} \frac{T_e}{2T_i} + \frac{3\kappa U(\frac{5}{2},\frac{5}{2}-\kappa;\alpha^2\kappa)}{2U(\frac{3}{2},\frac{3}{2}-\kappa;\alpha^2\kappa)}\right)^{1/2}, \tag{30}$$

$$\zeta \equiv \frac{\omega_r}{kv_{Te}} = \sqrt{\frac{m_e}{2m_i}}\left(\frac{1}{k^2 \lambda_{De}^2 + \frac{U(\frac{1}{2},\frac{1}{2}-\kappa;\alpha^2\kappa)}{\kappa U(\frac{3}{2},\frac{3}{2}-\kappa;\alpha^2\kappa)}} + \frac{3\kappa U(\frac{5}{2},\frac{5}{2}-\kappa;\alpha^2\kappa)}{U(\frac{3}{2},\frac{3}{2}-\kappa;\alpha^2\kappa)}\frac{T_i}{T_e}\right)^{1/2}. \tag{31}$$

When we take $\alpha = 0$, this expressions (16) and (29) can correctly recover to those of the plasma with the $q$-distribution (or the $\kappa$-distribution) in nonextensive statistics [8].

## 3. Numerical analyses

In this section, we make numerical analyses of the dispersion relation in Eq. (24) and the damping rate in Eq. (29) to show the effect of the parameter $\alpha$ on the ion acoustic waves. To do the numerical analyses conveniently, we may write Eq.(24) as

$$\frac{\omega_r^2}{\omega_{pi}^2} = k^2 \lambda_{De}^2 \left(\frac{1}{k^2 \lambda_{De}^2 + \frac{U(\frac{1}{2},\frac{1}{2}-\kappa;\alpha^2\kappa)}{\kappa U(\frac{3}{2},\frac{3}{2}-\kappa;\alpha^2\kappa)}} + 3\kappa \frac{T_i}{T_e}\frac{U(\frac{5}{2},\frac{5}{2}-\kappa;\alpha^2\kappa)}{U(\frac{3}{2},\frac{3}{2}-\kappa;\alpha^2\kappa)}\right), \tag{32}$$

where we have used $\lambda_{De}^2 \omega_{pi}^2 = k_B T_e / m_i$.

Based on Eq.(32) and Eq.(29), the numerical analyses of the dispersion relation and damping rate are made respectively as a function of the wave number $k\lambda_{De}$ for several different $\alpha$-parameter, where the basic plasma physical quantities are chosen as $m_i/m_e = 1837$ and $T_i/T_e = 0.01$.

In Fig.1, based on the dispersion relation (32), the wave frequency ($\omega_r/\omega_{pi}$) of ion-acoustic waves is showed as a function of the wave number $k\lambda_{De}$ and for different values of the $\alpha$-parameter, where Fig.1(a) is for the parameter $\kappa = 2$, and the line with $\alpha=0$ is corresponding to the case of the plasma with the kappa-distribution for $\kappa = 2$; Fig.1(b) is for the parameter $\kappa = 1$ (as an example of $\kappa < 3/2$) and $\alpha \neq 0$.

Fig.1 shows that with increase of the wave number $k\lambda_{De}$, the wave frequency ($\omega_r/\omega_{pi}$) will increase monotonically and will decrease as the $\alpha$-parameter increases. Therefore, the wave frequency ($\omega_r/\omega_{pi}$) of ion-acoustic waves in the plasma with the regularized kappa-distribution (12) is generally less than that with the kappa-distribution (15). And as expected, Fig.1(b) also shows that the ion-acoustic waves exist in the plasma with the regularized kappa-distribution if $\kappa < 3/2$.

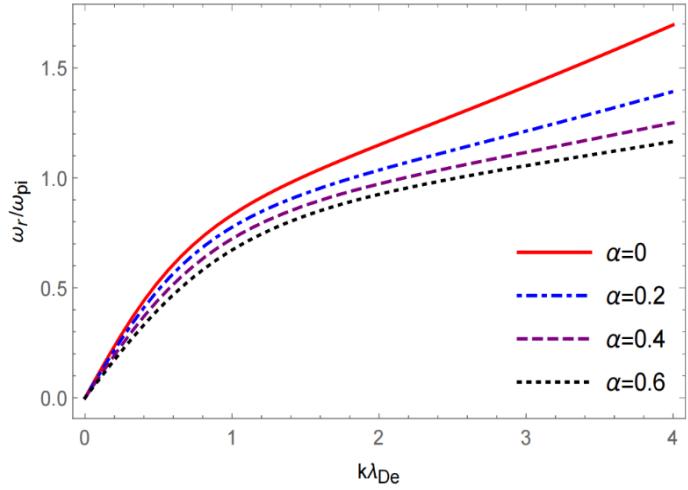

(a) For $\kappa=2$

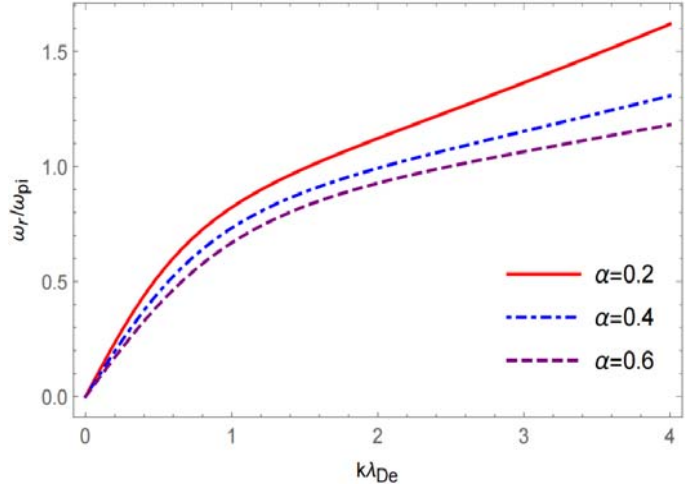

(b) For $\kappa=1$ and $\alpha \neq 0$.

**Figure 1.** The dispersion relation for different values of the $\alpha$-parameter.

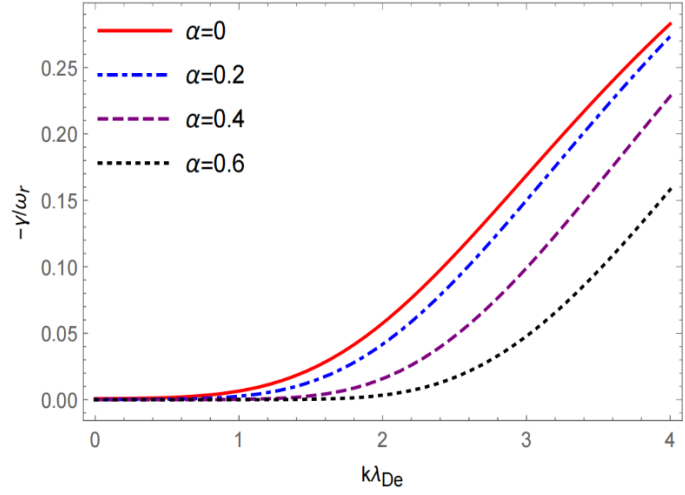

(a) For $\kappa = 4$

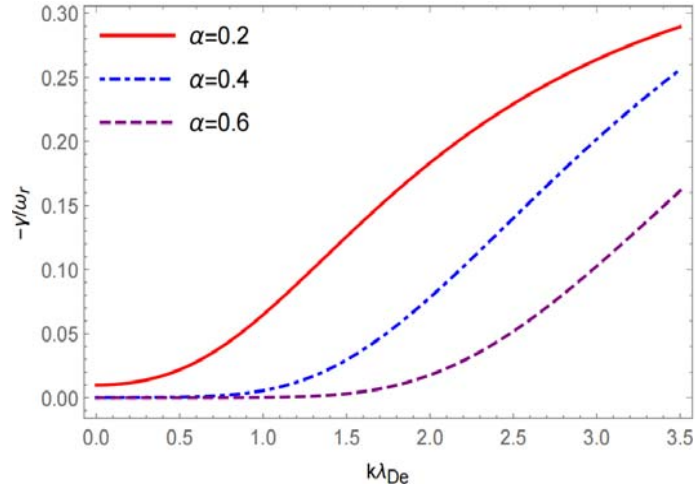

(b) For $\kappa = 1$ and $\alpha \neq 0$

**Figure 2.** The damping rate for different values of the $\alpha$-parameter.

In Fig.2, based on Eq.(29), the damping rate ($-\gamma / \omega_r$) of ion-acoustic waves is also showed as a function of the wave number $k\lambda_{De}$ and for four different values of the $\alpha$-parameter, where Fig.2(a) is for the parameter $\kappa = 4$, and the line with $\alpha = 0$ is corresponding to the case of the plasma with the kappa-distribution for $\kappa = 4$. Fig.2 (b) is for the parameter $\kappa = 1$ (as an example of $\kappa < 3/2$) and $\alpha \neq 0$.

Fig.2 shows that with increase of the wave number $k\lambda_{De}$, the damping rate ($-\gamma / \omega_r$) will increase monotonically and will also decrease as the $\alpha$-parameter increases. Therefore, the damping rate ($-\gamma / \omega_r$) of ion-acoustic waves in the plasma with the regularized kappa-distribution (12) is also generally less than that with the kappa-distribution (15). And as expected, Fig.2 (b) also shows that the damping rate of ion-acoustic waves exist in the plasma with the regularized kappa-distribution if $\kappa < 3/2$.

## 4. Conclusion

In conclusion, we have studied the dispersion relation and the damping rate of the ion-acoustic waves in the plasma with the so-called regularized kappa-distribution given by Eq.(12), an optimized kappa-distribution which is not only suitable for the parameter $\kappa > 3/2$, but also suitable for $0 < \kappa < 3/2$. We have derived the dispersion relation and the damping rate, which are given by Eq.(15) and Eq.(17), respectively. We find that the dispersion relation Eq.(15) and the damping rate Eq.(17) both depend significantly on the parameters $\alpha$ and $\kappa$.

The numerical analyses showed that the wave frequency ($\omega_r/\omega_{pi}$) and the damping rate ($-\gamma/\omega_r$) of ion-acoustic waves in the plasma with the regularized kappa-distribution are both generally less than those in the plasma with the kappa-distribution, and as expected, the ion-acoustic waves and their damping rate exist in the plasma with the regularized kappa-distribution if $\kappa < 3/2$.

**Appendix**

For ions ($\xi_i \gg 1$), one has that

$$\int_{-\infty}^{\infty} dx \left(1+\frac{x^2}{\kappa}\right)^{-\kappa-1} \frac{\exp(-\alpha^2 x^2)}{\xi_i - x} = \int_{-\infty}^{\infty} \exp(-\alpha^2 x^2)\left(1+\frac{x^2}{\kappa}\right)^{-\kappa-1}\left(\frac{1}{\xi_i}+\frac{x}{\xi_i^2}+\frac{x^2}{\xi_i^3}+\frac{x^3}{\xi_i^4}\right)dx, \quad (A.1)$$

where

$$\int_{-\infty}^{\infty} dx \left(1+\frac{x^2}{\kappa}\right)^{-\kappa-1} \frac{1}{\xi_i}\exp(-\alpha^2 x^2) = \frac{1}{\xi_i}\sqrt{\pi\kappa}\, U(\tfrac{1}{2},\tfrac{1}{2}-\kappa;\alpha^2\kappa),$$

$$\int_{-\infty}^{\infty} dx \left(1+\frac{x^2}{\kappa}\right)^{-\kappa-1} \frac{x^2}{\xi_i^3}\exp(-\alpha^2 x^2) = \frac{1}{2\xi_i^3}\sqrt{\pi}\,\kappa^{3/2} U(\tfrac{3}{2},\tfrac{3}{2}-\kappa;\alpha^2\kappa),$$

$$\int_{-\infty}^{\infty} dx \left(1+\frac{x^2}{\kappa}\right)^{-\kappa-1} \frac{x^4}{\xi_i^5}\exp(-\alpha^2 x^2) = \frac{3}{4\xi_i^5}\sqrt{\pi}\,\kappa^{5/2} U(\tfrac{5}{2},\tfrac{5}{2}-\kappa;\alpha^2\kappa).$$

For electrons ($\xi_e \ll 1$), one lets $x - \xi_e = \eta$ and

$$\int_{-\infty}^{\infty} dx \left(1+\frac{x^2}{\kappa}\right)^{-\kappa-1} \frac{\exp(-\alpha^2 x^2)}{\xi_e - x} = -\int_{-\infty}^{\infty} \frac{d\eta}{\eta}\left(1+\frac{(\xi_e+\eta)^2}{\kappa}\right)^{-\kappa-1}\exp(-\alpha^2(\xi_e+\eta)^2)$$

$$= -\int_{-\infty}^{\infty} \frac{d\eta}{\eta}\left(1+\frac{\xi_e^2+2\xi_e\eta+\eta^2}{\kappa}\right)^{-\kappa-1}\exp(-\alpha^2(\xi_e^2+2\xi_e\eta+\eta^2))$$

$$\approx -\int_{-\infty}^{\infty} \frac{d\eta}{\eta}\left(1+\frac{\eta^2}{\kappa}\right)^{-\kappa-1}\exp(-\alpha^2\eta^2) = 0. \quad (A.2)$$

Eq.(19) becomes

$$D_{\alpha,\kappa}(\omega,k) = 1 + \frac{2U(\tfrac{1}{2},\tfrac{1}{2}-\kappa;\alpha^2\kappa)}{\kappa U(\tfrac{3}{2},\tfrac{3}{2}-\kappa;\alpha^2\kappa)}\sum_s \frac{\omega_{ps}^2}{k^2}\frac{1}{v_{Ts}^2}\left[1-\frac{\xi_s/\sqrt{\pi\kappa}}{U(\tfrac{1}{2},\tfrac{1}{2}-\kappa;\alpha^2\kappa)}\mathrm{Pr}\int_{-\infty}^{\infty}dx\left(1+\frac{x^2}{\kappa}\right)^{-\kappa-1}\frac{\exp(-\alpha^2 x^2)}{\xi_s - x}\right]$$

$$= 1 - \frac{\omega_{pi}^2}{k^2 v_{Ti}^2}\frac{1}{\xi_i^2}\left(1+\frac{3\kappa}{2\xi_i^2}\frac{U(\tfrac{5}{2},\tfrac{5}{2}-\kappa;\alpha^2\kappa)}{U(\tfrac{3}{2},\tfrac{3}{2}-\kappa;\alpha^2\kappa)}\right) + \frac{2\omega_{pe}^2}{\kappa k^2 v_{Te}^2}\frac{U(\tfrac{1}{2},\tfrac{1}{2}-\kappa;\alpha^2\kappa)}{U(\tfrac{3}{2},\tfrac{3}{2}-\kappa;\alpha^2\kappa)}$$

$$+ i\sqrt{\frac{\pi}{\kappa^3}}\frac{2\omega}{U(\tfrac{3}{2},\tfrac{3}{2}-\kappa;\alpha^2\kappa)}\sum_s \frac{\omega_{ps}^2}{k^3 v_{Ts}^3}\left(1+\frac{\omega^2}{\kappa k^2 v_{Ts}^2}\right)^{-\kappa-1}\exp\left(\frac{-\alpha^2\omega^2}{k^2 v_{Ts}^2}\right). \quad (A.3)$$

Substituting $\xi_i = \omega/kv_{Ti}$ into Eq.(A.1), one obtains Eq.(20).

**Acknowledgements**

This work is supported by the National Natural science foundation of China under Grant No. 11775156.